\documentclass[a4paper,12pt]{article}
\usepackage{amsmath,amssymb}
\usepackage{graphicx,color}
\usepackage{mathrsfs}
\usepackage{cite}

  \makeatletter
  \@addtoreset{equation}{section}
  \makeatother
\usepackage[
      colorlinks=true,
      linkcolor=blue,
      urlcolor=blue,
      filecolor=black,
      citecolor=blue,
      pdfstartview=FitV,
      pdftitle={},
        pdfauthor={Ryotaku Suzuki, Shinya Tomizawa},
        pdfsubject={},
        pdfkeywords={},
        pdfpagemode=None,
        bookmarksopen=true,
      ]{hyperref}
\usepackage{caption}

\newcommand{\fr}[1]{\frac{1}{#1}}

\newcommand{\ord}[1]{{\mathcal O}\left(#1\right)}

\newcommand{\cA}{{\mathcal A}}

\newcommand{\nonum}{\nonumber\\ }

\newcommand{\sR}{{\sf R}}

\newcommand{\cout}[1]{}

\newcommand{\hgfunc}[2]{ { \, {}_{#1}  F  {}_{#2} } }

\usepackage{ulem}

\numberwithin{equation}{section}

\begin{document}

\begin{titlepage}
\rightline{TTI-MATHPHYS-7}

\vskip 2cm
\vglue 2cm
\centerline{\LARGE \bf Squashed black holes at large $D$}

\vskip 1.6 cm
\centerline{\bf Ryotaku Suzuki and Shinya Tomizawa}
\vskip 0.5cm

\centerline{\small Mathematical Physics Laboratory, Toyota Technological Institute}
\centerline{\small Hisakata 2-12-1, Nagoya 468-8511, Japan}
\smallskip
\vskip 0.5cm
\centerline{\small\tt sryotaku@toyota-ti.ac.jp, tomizawa@toyota-ti.ac.jp}
\vskip 1cm

\centerline{\bf Abstract} \vskip 0.2cm \noindent
\noindent
Using the large $D$ effective theory approach, we construct a static solution of non-extremal and squashed black holes with/without an electric charge, which describes a spherical black hole in a Kaluza-Klein spacetime with a compactified dimension.
The asymptotic background with a compactified dimension and near-horizon geometry are analytically solved by the $1/D$ expansion. Particularly, our work demonstrates that the large $D$ limit can be applied to solve the non-trivial background with a compactified direction, which leads to a first-order flow equation. Moreover, we show that the extremal limit consistently reproduces the known extremal result.

\end{titlepage}
\pagestyle{empty}
\small
\addtocontents{toc}{\protect\setcounter{tocdepth}{2}}
{
	\hypersetup{linkcolor=black,linktoc=all}
	\tableofcontents
}
\normalsize
\newpage

\pagestyle{plain}
\setcounter{page}{1}

\section{Introduction}

So far, higher-dimensional black holes have played  important roles in understanding basic properties of fundamental theories, such as string theory. 
A number of interesting solutions of such higher-dimensional black holes have been found and shown us  that they have much richer structure of their solution space than that of four-dimensional black holes~\cite{Emparan:2008eg}. 
However, since our observable world is macroscopically four-dimensional, extra dimensions have to be compactified in realistic spacetime models~\cite{Kaluza:1921tu,Klein:1926tv}. 
Therefore, from this point of view, it is of great importance to consider higher-dimensional Kaluza-Klein black holes, which look like a four-dimensional spacetime at least at large distances. 
One hopes that the studies on such Kaluza-Klein black holes may also give us some insights into the major open problem of how to compactify and stabilize extra dimensions in string theory. 
%
The simplest example of a five-dimensional Kaluza-Klein black hole is a black-string, which is a direct product of a four-dimensional black hole and a circle. 

\medskip
A more non-trivial class of Kaluza-Klein black holes is given by
{\sl squashed} Kaluza-Klein (SqKK) black holes which are obtained by applying the deformation of squashing to five-dimensional black holes. 
For instance, the basic idea is to view the $S^3$ section, $d\Omega_3^2$, of a five-dimensional Schwarzschild  black hole spacetime as a fiber bundle of $S^1$ over the $S^2$ base space $d\Omega_2^2$, as follows,
\begin{eqnarray*}
d\Omega_3^2&=&\frac{1}{4}[(d\psi+\cos\theta d\phi)^2+d\Omega_2^2], 
\end{eqnarray*}
and then perform such a deformation that changes the ratio of the radii of the fiber $S^1$ and the $S^2$ base, as
\begin{eqnarray*}
ds^2=-\left(1-\frac{m}{r^2}\right)dt^2+\left(1-\frac{m}{r^2} \right)^{-1}k^2(r)dr^2+\frac{r^2}{4}[(d\psi+\cos\theta d\phi)^2+k(r)d\Omega_2^2],
\end{eqnarray*}
where $k(r)$ is called squashing function, which is determined by the Einstein equation. 
After this squashing deformation, the resultant spacetime asymptotically looks like an $S^1$ fiber bundle over a base space of a four-dimensional  flat spacetime  at large distances, whereas it looks like a five-dimensional black hole near the horizon. 
The basic structure of SqKK black holes can in fact be seen in the much earlier works of refs.~\cite{Dobiasch:1981vh,Ishihara:2005dp}, whose solutions asymptote to an $S^1$-bundle over a four-dimensional  flat
spacetime, as studied in~\cite{Gibbons:1985ac}. 
Some further generalizations of SqKK black
holes has been made subsequently~\cite{Nakagawa:2008rm,Tomizawa:2008hw,Tomizawa:2008rh}. 

 \medskip
Moreover, the accumulation of this type of SqKK black hole solutions also motivates us to generalize to higher dimensions.  
For example, in ref.~\cite{Tatsuoka:2011tx}, odd-dimensional extremal charged black hole solutions with a compactified dimension were obtained  by squashing $S^{2n+1}$, which can be viewed as an $S^1$ fiber bundle over $CP^n$ base space, 
\begin{eqnarray*}
d\Omega_{2n+1}^2=(d\phi+{\cal A}_n)^2+d\Sigma_n^2,
\end{eqnarray*}
in terms of a metric $d\Sigma_n^2$ and a K\"ahler potential on $CP^n$,
then the ratio of the radii of the fiber $S^1$ and the base $CP^n$ is changed so that the ratio is to same extent on the horizon but diverges at infinity. 
The resultant spacetime looks like a black hole with $S^{2n+1}$ topology near the horizon but has a compact spatial direction of $\partial/\partial\phi$ at infinity.

\medskip
To explore a vast variety of higher dimensional black holes, the large dimension limit, or large $D$ limit, of gravitational theories provides a versatile analytic approach~\cite{Emparan:2013moa,Emparan:2020vyfinr}.
The basic feature of the large $D$ limit, the localization of the gravity,
confines the black hole dynamics within the thin layer of $\ord{1/D}$-thickness along the horizon, to form an effective theory living on the horizon surface~\cite{Emparan:2015hwa,Bhattacharyya:2015dva,Bhattacharyya:2015fdk}. This large $D$ effective theory approach facilitates the search for more general, less symmetric solutions. So far, this approach has been applied to study various types of black holes such as the black brane instability and related non-uniform branches~\cite{Emparan:2015hwa,Suzuki:2015axa,Emparan:2015gva,Emparan:2016sjk,Rozali:2016yhw,Emparan:2018bmi}, rotating compact black holes~ 
\cite{Tanabe:2015hda,Tanabe:2016opw,Chen:2017wpf,Mandlik:2018wnw} and other solutions in more complicated setups~\cite{Iizuka:2018zgt,Herzog:2017qwp}.
The instability, deformation and interaction of compact black holes can be systematically studied by the blob approximation, in which a compact black hole is identified as a Gaussian lump, or black blob, on the black brane effective theory~\cite{Andrade:2018rcx,Andrade:2018nsz,Andrade:2018yqu,Andrade:2019edf,Licht:2020odx,Andrade:2020ilm,Suzuki:2020kpx}.
The large $D$ effective theory is also applicable to Gauss-Bonnet black holes~\cite{Chen:2017rxa,Chen:2018vbv}.

 \medskip
In this paper, we use the technique of the large $D$ effective theory to find black hole solutions in the odd-dimensional SqKK background in Einstein and Einstein-Maxwell theories.
 Specifically, we focus on static and non-extremal black holes with  $S^{2n+1}$ topology, which can be regarded as an $S^1$ fiber bundle over $CP^n$ base space. Therefore, this is straightforward extensions of the five-dimensional black holes~\cite{Ishihara:2005dp} to a higher-dimensional case as well as of the $(2n+3)$-dimensional black holes~\cite{Tatsuoka:2011tx} to a non-extremal case.
As a novel feature, instead of imposing
simple backgrounds such as asymptotically flat or (A)dS,
we solve the non-trivial squashed background, more specifically generalized $(2n+2)$-dimensional Taub-NUT space background,
 by using the $1/D$ expansion besides the near-horizon analysis.\footnote{Using the large $D$ terminology, we solve the decoupled sector in both near horizon and asymptotic region.} This demonstrates that the large $D$ limit can also be useful in finding a class of
Kaluza-Klein spacetimes having non-trivial bundles
with/without a horizon. The charged solutions are also studied in the same analysis. 
 We can find that
 the extremal limit is consistent with the large $D$ limit of the known extremal solution~\cite{Tatsuoka:2011tx}.
 
 \medskip
This paper is organized as follows.
In section~\ref{sec:sqbg}, we start by revisiting the squashed background at large $D$. In section~\ref{sec:sqkk}, the near horizon geometry of neutral squashed black holes are solved in $1/D$-expansion. The resulting physical quantities are shown in section~\ref{sec:phys}. The section~\ref{sec:sqkk-Q} repeats the similar analysis for charged black holes. The extremal limit is also discussed. We summarize our result in section~\ref{sec:sum}.
We attached an auxiliary {\it Mathematica} notebook file in the supplementary material to present lengthy metric solutions in $1/D$-expansion.

\section{Squashed background}\label{sec:sqbg}
First, using the large $D$ limit, we reconstruct the squashed background with a compact $S^1$ direction, i.e., the $(2n+2)$-dimensional Eulidean Taub-NUT space which was studied in~ref.\cite{Tatsuoka:2011tx}, 
\begin{align}
ds_{2n+2}^2 = \frac{dr^2}{F(r)} + L ^2 F(r) (d\phi+\cA_n)^2+r(r+2L)d\Sigma_n^2, \label{eq:TN}
\end{align}
where $d\Sigma_n^2=\gamma_{ij}d\sigma^id\sigma^j$ is the $CP^n$-metric with the curvature $\hat{R}_{ij}=2(n+1)\gamma_{ij}$ and $\cA_n$ is the K\"{a}hler potential on $CP^n$. The $S^1$ direction is identified with $\phi \sim \phi+2\pi$.\footnote{Note that our convention of the $CP^n$ metric is different from ref.~\cite{Tatsuoka:2011tx}, in which the $S^1$ direction is identified with $\phi \sim \phi + 4(n+1)\pi$. We follow the construction in ref.~\cite{Hoxha:2000jf}.} Since the large dimension now owes to $CP^n$, we consider the large $n$ limit in the following.

The Ricci-flat condition of this geometry is given by
\begin{align}
(2 L^2 n + L ( 4 n-2) r + (2 n-1) r^2) F(r) + r (2 L + r) ( (L + r) F'(r)-2 (1 + n))=0.
\label{eq:strictFr-eq}
\end{align}
The solution regular at $r=0$ is obtained as
\begin{align}
F(r)=\frac{2r}{L}\left(1+\frac{r}{L}\right)\left(1+\frac{r}{2L}\right)^{-n}F_1\left(n+1,2,-n,n+2;-\frac{r}{L},-\frac{r}{2L}\right),
\label{eq:strictFr}
\end{align}
where $F_1$ is the Appell's double hypergeometric function which reduces to polynomial for an integer $n$.
Now, instead of direct integration, we observe the large $n$ limit of eq.~(\ref{eq:strictFr-eq})
\begin{align}
2n( L+r)^2 F(r) -2n r(2L+r) + \ord{n^0} = 0.
\end{align}
Interestingly, the derivative term goes to the sub-leading order, and
then the leading-order solution for $F(r)$ is determined by merely solving an algebraic equation. Solving eq.~(\ref{eq:strictFr-eq}) order by order in the $1/n$ expansion, we obtain
\begin{align}
F(r) = \frac{r(r+2L)}{(r+L)^2}+\frac{3r^2(r+2L)^2}{2n(r+L)^4}+\frac{3 r^2 (2 L+r)^2 \left(-4 L^2+2 L r+r^2\right)}{4n^2 (L+r)^6}+\cdots,
\label{eq:largedex-Fr}
\end{align}
which correctly approximates eq.~(\ref{eq:strictFr}) at large $n$.
Therefore, we can obtain the squashed background at the large $D$ limit as
\begin{align}
ds_{2n+2}^2 \simeq \frac{(r+L)^2}{r(r+2L)}dr^2 + \frac{L^2 r(r+2L)}{(r+L)^2} (d\phi+\cA_n)^2+r(r+2L)d\Sigma_n^2.\label{eq:squashedbk_revist_largeDsol}
\end{align}

\subsection{Squashed background at large $D$}
In the above analysis, we have seen that the large $D$ limit correctly reproduces the squashing behavior of the known result~\cite{Tatsuoka:2011tx} in the $1/n$-expansion~(\ref{eq:largedex-Fr}).
Here, we show that the spacetime with an $S^1$ fiber bundle over $CP^n$
 admits the squashing deformation at large $n$ even if we use a more general metric ansatz rather than eq.~(\ref{eq:TN}).
Let us start from the following general ansatz with an $S^1$ fiber bundle over $CP^n$,
\begin{align}
ds^2 = Fdr^2 + G_{ab}dx^a dx^b + 2G_{a\phi} dx^a (d\phi+\cA_n)+G_{\phi\phi} (d\phi+\cA_n)^2+r^2 d\Sigma_n^2, 
\end{align}
where $d\Sigma_n=\gamma_{ij}d\sigma^i d\sigma^j$ and $\cA_n$ are the same $CP^n$ metric and  K\"{a}hler potential,~respectively,  and the coordinates $x^a$ may include another spatial coordinates as well as a time coordinate $t$. 
Here, we suppose that the metric components are function of $r$ and $x^a$. 
It is shown that by taking the limit $n\to\infty$, the Einstein equation in  Appendix~\ref{sec:A} reduces to
\begin{align}
& F = 1 + \ord{n^{-1}},\\
&\partial_r G_{AB} = \frac{2G_{A\phi}G_{B\phi}}{r^3}+\ord{n^{-1}},\label{eq:eqGAB}
\end{align}
where $A,B=a,\phi$.
For $G_{\phi\phi}$, this is immediately solved as
\begin{align}
 G_{\phi\phi} = \frac{L^2r^2}{L^2+r^2},
\end{align}
where $L$ is an integration constant, which causes to squash the background and compactify the fiber direction $\partial/\partial\phi$.
The other remaining components are solved as
\begin{align}
 &G_{a\phi} = \frac{c_a L^2 r^2}{r^2+L^2},\\
 &G_{ab} = H_{ab} - \frac{c_a c_b L^4}{r^2+L^2},
\end{align}
where $c_a$ and $H_{ab}$ are independent of $r$, and $c_a$ gives the boost $d\phi \to d\phi + c_a dx^a$ at $r \to \infty$.

One can easily confirm that this recovers eq.~(\ref{eq:squashedbk_revist_largeDsol}) by switching the radial coordinate.
In the resultant spacetime, the size of $S^1$  is finite and much smaller than the size of $2\pi L$ at infinity, so that the $S^1$ direction is compactified.
Moreover, we note that the limit $L\to\infty$ recovers $SO(2n+1)$ symmetry.
It is also worth noting that, in the leading order, $L$ is rather an integration function which can depend on other directions $x^a$, as eq.~(\ref{eq:eqGAB}) only solves the radial dependence. The sub-leading analysis will impose constraints on the form of $L(x)$.

\medskip
One should note that in the conventional large $D$ analysis, the background is usually put by hand, and the main focus has been always on the horizon.
Here, we have demonstrated that the large $D$ limit also provides an easy way to construct a squashed horizonless background spacetime with a compact dimension, which is determined by solving the geometric flow-like equation~(\ref{eq:eqGAB}).

\subsection{Asymptotic behavior in SqKK background}
Next, we elaborate the SqKK background at large $D$, but use the following Bondi ansatz, which is convenient for the later near horizon analysis
\begin{align}
ds^2 =  -dt^2+h_{rr}(r) dr^2 +h_{\phi\phi}(r)(d\phi+\cA_n)^2+r^2 d\Sigma_n^2.
\label{eq:sqKK-bg-flat}
\end{align}
Here we repeat the analysis in the previous section up to higher order in $1/n$,
\footnote{Although the current interest is the $1/n$ expanded behavior, we also have the strict solution by
\begin{equation}
h_{rr} = \fr{F(r)},\quad h_{\phi\phi} = \frac{\tilde{L}^2r^2F(r)}{r^2+\tilde{L}^2}
\end{equation}
where
\begin{equation}
 F(r) = \frac{2(n+1)}{2n-1}\hgfunc{2}{1} \left[1,\frac{3}{2},\frac{3}{2}-n,\frac{\tilde{L}^2}{\tilde{L}^2+r^2}\right]+\frac{C(r^2+\tilde{L}^2)^{3/2}}{r^{2n+2}},\quad \tilde{L} = \sqrt{\frac{2n-1}{2n+2}}L.
\end{equation}
The constant $C$ must vanish for the regularity.}
\begin{equation}
 h_{\mu\nu} = h^{[0]}_{\mu\nu} + \frac{1}{n} h_{\mu\nu}^{[1]}+\cdots,
\end{equation}
which gives 
\begin{subequations}\label{eq:squashedbk-sol-bondi-LD}
\begin{align}
& h_{rr}(r) = 1 - \frac{3r^2}{2n(L^2+r^2)}+\frac{3 r^2 \left(L^2+2 r^2\right)}{4 n^2 \left(L^2+r^2\right)^2}+\ord{n^{-3}},\\
& h_{\phi\phi}(r) = \frac{L^2 r^2}{L^2+r^2} + 0\times n^{-1}-\frac{3 r^4 L^4}{2n^2(L^2+r^2)^3} + \ord{n^{-3}}.
\end{align}
\end{subequations}
Under this ansatz, the Einstein equation can be also solved by expanding the metric from $r=\infty$ in small $L/r$, and we have
\begin{subequations}\label{eq:squashedbk-sol-bondi-as}
\begin{align}
&h_{rr}(r) = \frac{2n-1}{2n+2}+\frac{3 (2 n-1)^2}{4 (n+1)^2 (2 n-3)}\frac{L^2}{r^2}+\ord{\frac{L^4}{r^4}},\\
&h_{\phi\phi}(r) = L^2 \left(1-\frac{n (2 n-1)}{(n+1) (2 n-3)}\frac{L^2}{r^2}+\ord{\frac{L^4}{r^4}}\right).
\end{align}
\end{subequations}
These two expansions are consistently matched by the double expansion with $1/n$ and $L/r$.
One can easily see from eq.~(\ref{eq:squashedbk-sol-bondi-as}) that  the metric~(\ref{eq:sqKK-bg-flat}) asymptotes to the Kaluza-Klein spacetime compactified in  the $S^1$ direction with the radius $2\pi L$ for $r\to\infty$
\begin{align}
 ds^2 \simeq -dt^2+\frac{2n-1}{2n+2} dr^2 + L ^2 (d\phi+\cA_n)^2+r^2 d\Sigma_n^2.
\end{align}

\paragraph{Squashed perturbation}
Given the squashed background~(\ref{eq:sqKK-bg-flat}), it is natural to expect the perturbative behavior is affected by the squashing effect.
Let us assume the linear perturbation to the background~(\ref{eq:sqKK-bg-flat}) as,
\begin{align}
g_{tt} = -1+\frac{a(r)}{r^{2n-1}},\quad g_{rr} \simeq h_{rr}(r)\left(1+\frac{b(r)}{r^{2n-1}}\right) ,\quad g_{\phi\phi} \simeq h_{\phi\phi}(r)\left(1+\frac{c(r)}{r^{2n-1}}\right),
\label{eq:squashed-monopoles-ansatz}
\end{align}
where the typical falling factor $r^{-2n+1}$ is separated in advance, so that $a(r),b(r)$ and $c(r)$ remains finite functions of $r$ at $n\to \infty$.
Then, the linearized equation is solved by $1/n$-expansion,
\begin{align}
& a(r) = \alpha \sqrt{1+\frac{L^2}{r^2} }+\ord{n^{-1}},\quad
 b(r) =\alpha\sqrt{1+\frac{L^2}{r^2} }-\beta \left(1+\frac{L^2}{r^2}\right)^{3/2}  +\ord{n^{-1}},\nonum
&  c(r) = \beta \left(1+\frac{L^2}{r^2}\right)^{3/2}-\frac{L^2 \left(2 \alpha  \left(L^2+2 r^2\right)+3 \beta 
   \left(L^2+r^2\right)\right)}{4 r^3 \sqrt{L^2+r^2}\,n}+\ord{n^{-2}},
 \label{eq:squashed-monopoles}
\end{align}
where $\alpha$ and $\beta$ are the integration constants and $c(r)$ is shown up to $\ord{n^{-1}}$ for the later use. 
The $\ord{(L/r)^0}$ terms at each order of $1/n$ can be absorbed into $\alpha$ and $\beta$ by fixing the integration constants, and hence one can identify them to the leading order terms in the asymptotic behavior at $r\to\infty$\footnote{One should not confuse with the linear perturbation~(\ref{eq:squashed-monopoles-ansatz}) in which $r$ is assumed finite but just $r^{-2n} \ll1$.}
\begin{equation}
 g_{tt} \simeq -1 + \frac{\alpha}{r^{2n-1}},\quad
  \frac{  g_{rr}}{h_{rr}} \simeq 1 + \frac{\alpha-\beta}{r^{2n-1}},\quad
  \frac{g_{\phi\phi}}{h_{\phi\phi}} \simeq 1 + \frac{\beta}{r^{2n-1}}.
\end{equation}
The parameters in eq.~(\ref{eq:squashed-monopoles}) are later matched with the near horizon solution.
Once the asymptotic behavior is determined, the ADM mass and tension are evaluated as follows~\cite{Harmark:2004ch},
\begin{align}
\textsc{Mass} =  \frac{\Omega_{2n+1}}{16 \pi G}\sqrt{\frac{2n+2}{2n-1}}L(2n \alpha-\beta)=\frac{n\Omega_{2n+1}}{8\pi G}{\cal M},
\end{align}
\begin{align}
\textsc{Tension}=\frac{\Omega_{2n+1}}{32\pi^2 G}\sqrt{\frac{2n+2}{2n-1}}(\alpha-2n \beta)=\frac{n\Omega_{2n+1}}{16\pi^2 G}{\cal T},
\end{align}
where $\Omega_{2n+1}=2\pi\, {\rm vol}(CP^n)$ is the volume of $S^{2n+1}$ and the normalized mass and tension is given by
\begin{equation}
{\cal M} =\sqrt{\frac{2n+2}{2n-1}}L\left( \alpha-\frac{\beta}{2n}\right),
\quad {\cal T} =\sqrt{\frac{2n+2}{2n-1}}\left(\frac{\alpha}{2n}- \beta\right).
\end{equation}
Similarly, the Komar mass is obtained as
\begin{equation}
\textsc{Komar Mass} = -\frac{2n+1}{32\pi G n} \int \nabla^\mu \xi^\nu dS_{\mu\nu} = \frac{n \Omega_{2n+1}}{8\pi G} {\cal M}_K,
\end{equation}
and
\begin{equation}
 {\cal M}_K =\frac{(2n+1)\sqrt{2(n+1)(2n-1)}}{4n^2} L \alpha.
\end{equation}

\section{Near-horizon analysis}\label{sec:sqkk}
In this section, we apply the standard near-horizon analysis at large $D$ to the squashed background. Bearing in mind the metric form of the background~(\ref{eq:sqKK-bg-flat}), we assume the following ansatz: 
\begin{align}
 ds^2 = - A(r) dt^2 + h_{rr}(r)\frac{B(r)}{A(r)}dr^2 + h_{\phi\phi}(r) H(r) (d\phi+\cA_n)^2+r^2d\Sigma_n^2,\label{eq:ansatz-nearH}
\end{align}
where $h_{rr}(r)$ and $h_{\phi\phi}(r)$ are the metric components for the squashed background, which are already solved in the $1/n$-expansion~(\ref{eq:squashedbk-sol-bondi-LD}). 
Since the metric has only $r$-dependence, we do not need to solve the large $D$ effective equation. Once we get the metric solution in the $1/n$ expansion, they trivially satisfies an effective equation, which must be solved in analysis for non-uniform black strings as in refs.~\cite{Emparan:2015hwa,Suzuki:2015axa}.
As is done in a usual analysis of the large $D$ effective theory, we introduce a new radial coordinate to resolve the thin near horizon region located around $r = r_0$ by
\begin{align}
 {\sf R} := (r/r_0)^{2n},
\end{align}
or inversely,
\begin{align}
 r = r_0 {\sf R}^\fr{2n} \simeq r_0 \left(1 + \fr{2n} \log {\sf R}\right).
\end{align}
The horizon scale is set to $r_0=1$ by rescaling $L$.
The background solution $h_{rr}(r)$ and $h_{\phi\phi}(r)$ should be expanded in $1/n$ again with the new coordinate.
The metric solution is obtained by expanding in $1/n$ as functions of ${\sf R}$,
\begin{align}
 A = \sum_{i=0}^\infty \frac{A_i({\sf R})}{n^i},\quad B = 1 + \fr{n}\sum_{i=0}^\infty \frac{B_i({\sf R})}{n^{i}},\quad H = 1 +\fr{n} \sum_{i=0}^\infty \frac{H_i({\sf R})}{n^{i}}.
 \label{eq:nearH-ansatz-expansion}
\end{align} 
We also impose the asymptotic boundary conditions at ${\sf R} \to \infty$ on these functions so that the geometry at infinity behaves as the squashed background by
\begin{align}
 A= 1+\ord{{\sf R}^{-1}} ,\quad B=1+\ord{{\sf R}^{-1}},\quad H =1+\ord{{\sf R}^{-1}}. \label{eq:asymp}
\end{align}

\medskip
If we require the existence of a horizon at ${\sf R} = m$ (a certain positive constant), the leading function $A_0$ is determined as
\begin{eqnarray}
 A_0 = 1 - \frac{m}{\sf R}.
\end{eqnarray}
Moreover, from the requirement of regularity on the horizon and asymptotic boundary condition~(\ref{eq:asymp}), the other leading functions are obtained as
\begin{align}
 \quad B_0 = 0,\quad H_0=0.
\end{align}
Furthermore, we require that the metric functions should be regular on the horizon even at each higher order in $1/n$.
Choosing the integration constants included in $A_i\ (i=1,\ldots)$ appropriately, we can fix the horizon at the same position ${\sf R} = m$ in all orders of $1/n$.

\medskip

Matching with the asymptotic perturbation~(\ref{eq:squashed-monopoles}) for $1\ll {\sf R}\ll e^{2n}$, we can extract the asymptotic behavior at the leading order,
\begin{align}
 \alpha = \frac{m}{\sqrt{1+L^2}}+\ord{n^{-1}},\quad \beta = \ord{n^{-1}}.
\end{align}
One can see that the asymptotic behavior in the near-horizon region gets a dressing factor in the asymptotic region due to the squashing effect.\footnote{To take the limit $L\to\infty$, $\alpha$ and $\beta$ must be rescaled by $\alpha \to \alpha/L$ and $\beta \to \beta/L^3$ so that the asymptotic behavior~(\ref{eq:squashed-monopoles}) remains finite, in which the match reproduces non-squashed result $\alpha = m$.}
Since $H$ is solved up to $\ord{n^{-1}}$, the match with $c(r)$ in  eq.~(\ref{eq:squashed-monopoles}) determines
\begin{equation}
\beta = \frac{L^2(L^2+2)m}{2(1+L^2)^{5/2}n}+\ord{n^{-2}}.
\end{equation}
$B$ and $b$ match accordingly.

Continuing the analysis order by order, we obtained the metric solution up to NNNLO, part of which is shown in Appendix~\ref{app:sol-neutral} ( See the attached auxiliary file in the supplementary material for the full data ). 

\section{Physical Properties}\label{sec:phys}
Matching the near horizon solution with the asymptotic behavior at infinity, we can compute the ADM mass ${\cal M}$, Komar mass ${\cal M}_K$ and tension ${\cal T}$ in $1/n$-expansion up to the next-to-next-to-leading order (NNLO) in $1/n$, which are written as, respectively, 
\begin{align}
&{\cal M}=\frac{L m}{\sqrt{L^2+1}}\left[1+\frac{ 2 L^2-2 \log
   m+3}{4 \left(L^2+1\right) n}
   +\frac{ \left(12-24 L^2\right) \log ^2 m+8 \left(\pi ^2-9\right) L^2-36 \log
   m+9}{96 \left(L^2+1\right)^2 n^2}\right],\\
&{\cal T}=\frac{m}{2 \left(L^2+1\right)^{5/2}n}\left[1
 +\frac{ \left(8 L^2-2\right) \log m-2   L^2+3}{8 \left(L^2+1\right) n}\right.\nonum
   &\quad \left.   +\frac{12 \left(16 L^4-18 L^2+1\right) \log ^2m-12 \left(8 L^4-24   L^2+3\right) \log m+312 L^4+8 \left(\pi ^2-54\right) L^2+9}{192
   \left(L^2+1\right)^2 n^2} \right],
\end{align}
and 
\begin{align}
{\cal M}_K = \frac{Lm}{\sqrt{L^2+1}}\left[1+\frac{ 2 L^2-2 \log
   m+3}{4 \left(L^2+1\right) n}+\frac{\left(12-24 L^2\right) \log ^2 m+8 \left(\pi ^2-9\right) L^2-36 \log m-15}{96
   \left(L^2+1\right)^2 n^2}\right].
\end{align}
The surface gravity and horizon area are derived from the near horizon geometry
\begin{align}
\textsc{Surface\, gravity} = n\kappa,\quad
\textsc{Area} = \Omega_{2n+1} {\cal A}_H,
\end{align}
where the normalized quantities are given by
\begin{align}
 \kappa = 1+\frac{\frac{1}{4   \left(L^2+1\right)}-\frac{\log m}{2}}{n}
   + \frac{ 4( L^2+1)^2\log ^2m+4( L^2-1) \log m-4 L^2-9}{32 \left(L^2+1\right)^2 n^2},
\end{align}
and
\begin{align}
 {\cal A}_H =\frac{L m}{\sqrt{L^2+1}}\left[1+\frac{L^2  \log m}{2 \left(L^2+1\right)  n}+ \frac{L^2  \left(3 \left(L^2-2\right) \log ^2 m+2 \left(\pi ^2-9\right)\right)}{24
   \left(L^2+1\right)^{2} n^2}\right].
\end{align}
Assuming $L$ and $m$ as independent parameters, we can check the following relations \begin{align}
& d{\cal M} =  \kappa\, d{\cal A} _H+ {\cal T} dL,\\
&2n {\cal M} = (2n+1) \kappa {\cal A}_H+ {\cal T}L,\\
&{\cal M}_K = \frac{2n+1}{2n}\kappa {\cal A}_H ,
\end{align}
up to the given order of $1/n$.
Especially, the tension is proportional to the difference between ${\cal M}$ and ${\cal M}_K$
\begin{equation}
{\cal M}-{\cal M}_K = \frac{{\cal T}L}{2n},
\end{equation}
which was shown in ref.\cite{Kurita:2007hu} for a five-dimensional static and charged case.

\paragraph{Scaling invariant expression}
As we fixed the horizon scale $r_0=1$, we have the freedom to change the length scale of entire system.
One can check that, under the change of parameter 
\begin{align}
L \to C^{\frac{1}{2n}}L,\quad m \to C m,
\end{align}
 the above physical quantities must be subject to the following scaling laws up to the given order of $1/n$,
 \begin{equation}
  {\cal M} \to C {\cal M},\quad
  {\cal T} \to C^\frac{2n-1}{2n} {\cal T},\quad 
{\cal A}_H \to C^\frac{2n+1}{2n} {\cal A}_H,\quad \kappa \to C^{-\fr{2n}}\kappa.
 \end{equation}
 Thus, we can separate the scaling dependence from the scale invariance as
\begin{align}
{\cal M} = r_H^{2n} \widetilde{\cal M}(r_H/L) ,\
{\cal T} = r_H^{2n-1} \widetilde{\cal T}(r_H/L),\
{\cal A}_H = r_H^{2n+1} \widetilde{\cal A}_H(r_H/L),\
\kappa = r_H^{-1} \widetilde{\kappa}(r_H/L),
\end{align}
where $r_H := m^\fr{2n}$ is the radius of the $CP^n$ base on the horizon and the scale-invariant functions, $\widetilde{\cal M}$, $\widetilde{\cal T}$, $\widetilde{\kappa}$ and $\widetilde{\cal A}_H$, are given by
\begin{subequations}\label{eq:neutral-scaleinv}
\begin{align}
&\widetilde{\cal M}(x) = \frac{1}{\sqrt{x^2+1}}\left[1+\frac{2+3x^2}{4 (x^2+1) n}+\frac{8 \left(\pi ^2-9\right) x^2+9x^4}{96   \left(x^2+1\right){}^2 n^2}\right.\nonum
&\left.\hspace{4cm}+\, \frac{2 x^4 \left(20 \pi ^2-96 \zeta (3)-189\right)+48 x^2 (9-8 \zeta (3))+45 x^6}{384 n^3
   \left(x^2+1\right)^3}   \right],
\end{align}
\begin{align}
&\widetilde{\cal T}(x)= \frac{x^5}{2n(1+x^2)^{5/2}}\left[1+\frac{3x^2-2}{4 (x^2+1)n}
+\frac{312+8 \left(\pi ^2-54\right)   x^2+9x^4}{96 \left(x^2+1\right){}^2 n^2} \right.\nonum
&\left. +\, \frac{2 x^4 \left(20 \pi ^2-96 \zeta (3)-1215\right)+8 x^2 \left(4 \pi ^2-240 \zeta (3)+945\right)+45
   x^6+48 (32 \zeta (3)-49)}{384 n^3 \left(x^2+1\right)^3}  \right],
\end{align}
\begin{align}
\widetilde{\kappa}(x) =  1+\frac{x^2}{4 (x^2+1)n}-\frac{4 x^2+9}{32 \left(x^2+1\right){}^2 n^2}+\frac{x^2 \left(3 \left(9 x^4+36 x^2+56\right)-16 \pi ^2 \left(x^2+1\right)\right)}{384 n^3
   \left(x^2+1\right)^3},
\end{align}
\begin{align}
\widetilde{\cal A}_H(x) = \frac{1}{\sqrt{x^2+1}}\left[1+\frac{\left(\pi ^2-9\right) x^2}{12(x^2+1){}^2 n^2}+
   \frac{x^2 \left(x^2 \left(2 \pi ^2-12 \zeta (3)-9\right)-24 \zeta (3)+27\right)}{24 n^3
   \left(x^2+1\right)^3}
   \right],
\end{align}
\end{subequations}
where we have used the solution up to NNNLO to obtain the formula up to $\ord{n^{-3}}$.

\paragraph{Squashing function}
To measure the squashing effect on the horizon, it is convenient to calculate the squashing function by
\begin{equation}
 k_{\rm sq} := r^2/g_{\phi\phi}\bigr|_{{\sf R}=m},
\end{equation}
which is the ratio of the $CP^n$ radius to the $S^1$ radius.
By definition, $k_{\rm sq}$ is scale invariant, and hence fully expressed by the scale invariant function $k_{\rm sq}=\widetilde{k_{\rm sq}}(r_H/L)$, which is given by
\begin{align}
 \widetilde{k_{\rm sq}}(x)=1+x^2-\frac{(\pi ^2-9)x^2 }{6
   \left(x^2+1\right)n^2}+\frac{x^4 \left(12 \zeta (3)+9-2 \pi ^2\right)+3 x^2 (8 \zeta (3)-9)}{12 n^3 \left(x^2+1\right)^2},\label{eq:ksq-x}
   \end{align}
up to NNNLO. One can check the squashing is dissolved for $r_H/L \to 0$. It turns out the leading order terms alone approximate quite well for any $n\geq 1$,
\begin{align}
 \widetilde{k_{\rm sq}}(x) \approx 1+x^2- \frac{0.145}{n^2}\frac{x^2}{1+x^2}
 \approx 1+x^2.
\end{align}
This means that  the horizon must be oblate, namely,  the size of  $CP^n$  base is larger than the size of the $S^1$ fiber, as observed in $D=5$~\cite{Ishihara:2005dp}, due to compactified spatial infinity.


\section{Charged SqKK black holes}\label{sec:sqkk-Q}
Next, let us consider to find static, non-extremal and charged squashed black holes at large $D$ limit in D-dimensional Einstein-Maxwell theory, whose equation of motions are given by
the Einstein equation
\begin{equation}
R_{\mu\nu} - \fr{2} R g_{\mu\nu} = 2 \left(F_{\mu\alpha} F_{\nu}{}^\alpha - \fr{4}F^2 g_{\mu\nu}\right),
\end{equation}
and the Maxwell equation
\begin{equation}
\nabla_\mu F^{\mu\nu}=0,
\end{equation}
where $F_{\mu\nu} := \partial_\mu A_\mu - \partial_\nu A_\mu$. 
To this end, we assume the same ansatz as eq.~(\ref{eq:ansatz-nearH}) for the metric and the following form for the gauge field
\begin{equation}
 A_\mu dx^\mu = \Phi dt,
\end{equation}
where $\Phi=\Phi(r)$ corresponds to an electric potential.
The squashed perturbation and asymptotic behavior of the gauge field can be similarly solved to give
\begin{align}
& \Phi = \gamma \frac{\sqrt{r^2+L^2}}{r^{2n}} + \ord{n^{-1},r^{-4n}} \label{eq:phi-asym1}\\
 &\ \  = \frac{\gamma}{r^{2n-1}}+\ord{r^{-2n-1}}\label{eq:phi-asym2},
\end{align}
where $\gamma$ is the constant which is related to the normalized electric charge  as
\begin{equation}
Q = \frac{2\sqrt{(2n+2)(2n-1)}}{n} L\, \gamma, 
\end{equation}
and the physical electric charge is written in terms of $Q$ as
\begin{equation}
\textsc{Charge} = -\fr{8\pi G}\int F^{\mu\nu} dS_{\mu\nu} = \frac{n \Omega_{2n+1}}{8\pi G}Q.
\end{equation}

\paragraph{Near horizon analysis}
The near horizon analysis for the charged case is almost parallel to the neutral case performed in the previous sections, and we use the same near horizon coordinate ${\sf R} := r^{2n}$ and the same metric expansion~(\ref{eq:nearH-ansatz-expansion}).
Moreover, we assume that the electric potential is expanded in $1/n$ as 
\begin{equation}
 \Phi = \sum_{i=0}^\infty n^{-i}\Phi_i.
\end{equation}
If we impose the suitable  boundary conditions on the potential $\Phi$ such that (1) $\Phi\to 0$ at infinity ${\sf R} \to \infty$ and  (2) $\Phi $ is regular on the horizon, 
then we can obtain the leading order solution as
\begin{align}
& \Phi_0 = \frac{ \sqrt{\rho_+ \rho_-}}{\sqrt{2} {\sf R}},\quad
 A_0 = 1-\frac{\rho_++\rho_-}{{\sf R}}+\frac{\rho_+\rho_-}{{\sf R}^2},\nonum
 &H_0 = \frac{1}{1+L^2}\log \left(1-\frac{\rho_-}{\sf R}\right),\quad
  B_0 = -\frac{\rho_-}{(1+L^2)({\sf R}-\rho_-)}.\label{eq:charged-metricsol-LO}
\end{align}
We use the ambiguity in the integral constants $\rho_+,\ \rho_-$ in higher order, to fix the horizon position and the potential value on the horizon
\begin{equation}
 A\bigr|_{\sR = \rho_+}=0,\quad \Phi\bigr|_{\sR = \rho_+} = \sqrt{\frac{\rho_-}{2\rho_+}}.\label{eq:charge-parametrization}
\end{equation}
Matching the asymptotic behavior~(\ref{eq:squashed-monopoles}) and (\ref{eq:phi-asym1}) (up to leading order) leads to
\begin{align}
& \alpha = \frac{\rho_+\rho_-}{\sqrt{1+L^2}},\quad
\gamma = \sqrt{\frac{\rho_+\rho_-}{2(1+L^2)}}, \quad
\beta = \frac{\rho_+ L^2(L^2+2)+\rho_-(L^4+2L^2+2)}{2n(1+L^2)^{5/2}}.
\end{align}
Thus, we have solved  the equations of motion for the metric functions at large D up to NNLO and part of NNNLO to study the thermodynamics. In the appendix~\ref{app:sol-charged}, the metric solutions are given up to NLO.

\subsection{Physical quantities}
From the above results, we find that the thermodynamic variables are written as, up to $\ord{n^{-2}}$,
\begin{subequations}\label{eq:thermo-Q}
\begin{align}
& {\cal M}=\frac{L}{\sqrt{1+L^2}}\left[\rho_++\rho_-+\frac{\rho_+ \left(2 L^2+3\right)+\rho_--2 (\rho_++\rho-) \log (\rho_+-\rho_-)}{4 \left(L^2+1\right)n}
\right.\nonum
&\quad+\fr{n^2}\left(-\frac{\rho_-^2}{2(\rho_+-\rho_-)(1+L^2)}
+\frac{\left(4 \left(2 \pi ^2-27\right) L^2-75\right) \rho _-+\left(8 \left(\pi ^2-9\right)
   L^2+9\right) \rho _+}{96(1+L^2)^2}\right.\nonum
&\left.\left.\quad-\fr{8(1+L^2)^2}\left(\left(2 L^2-1\right) \left(\rho _-+\rho _+\right) \log ^2\left(\rho _+-\rho _-\right)
+\left(-2 L^2 \rho _-+\rho _-+3 \rho _+\right)\log \left(\rho _+-\rho _-\right) \right)\right)\right],\\
& Q = \frac{2L\sqrt{2\rho_+\rho_-}}{\sqrt{1+L^2}}
 \left[1+\frac{1-2 \log (\rho_+-\rho_-)}{4( L^2+1)n} +\fr{96n^2
   \left(L^2+1\right)^2}\left(-\frac{24\rho_-(1+L^2)}{\rho_+-\rho-}\right.\right.\nonum
&\left.\left. \qquad-12\left(2L^2-1\right)( \log ^2(\rho_+-\rho_-)- \log (\rho_+-\rho_-))+\left(8 \pi ^2-84\right) L^2-27\right)\right],\\
&{\cal T} = \frac{\rho_+-\rho_-}{2(1+L^2)^{5/2}n} \left[1+\fr{n}\left(\frac{\left(4 L^2-1\right) \rho _-+\left(3-2 L^2\right) \rho _+}{4   (\rho_+-\rho_-)\left(L^2+1\right)}-\frac{\left(4 L^2-1\right)  \log   \left(\rho _+-\rho _-\right)}{2 \left(L^2+1\right)}\right)\right],
\end{align}
\begin{align}
&{\cal A}_H = \frac{L \rho_+}{\sqrt{1+L^2}}\left[1+
\frac{1}{2n} \left(\log \rho _+ -\frac{\log   \left(\rho _+-\rho _-\right)}{L^2+1}\right)
+\right.\nonum
&+\fr{n^2}\left(\frac{
   \left(3-6 L^2\right) \log ^2\left(\rho _+-\rho _-\right)+2 \left(\pi
   ^2-9\right) L^2-6 \left(L^2+1\right) (\rho _-/\rho_+)}{24   \left(L^2+1\right)^2}\right.\nonum
&\left.\left.
   -\frac{ \log \left(\rho _+-\rho _-\right) \log
   \rho _+}{4 (1+L^2)}+\fr{8}\log ^2\rho _+-\frac{\rho _-^2}{4\rho_+ \left(L^2+1\right) \left(\rho _+-\rho _-\right)}\right)\right],\\
&\kappa = \frac{\rho_+-\rho_-}{\rho_+} + \frac{2 L^2 \rho _--2 \left(L^2+1\right) \left(\rho _+-\rho _-\right) \log \rho
   _++\rho _-+\rho _+}{4 \left(L^2+1\right) n \rho _+}\nonum
&+ \fr{n^2}\left(  \frac{\left(8 L^4+8 L^2-5\right) \rho _-+\left(4 L^2+9\right) \rho _+}{32   \left(L^2+1\right)^2 \rho _+}+\frac{L^2 \left(\rho _--\rho _+\right) \log
   \left(\rho _+-\rho _-\right)}{4 \left(L^2+1\right)^2 \rho _+}\right.\nonum
 &\left.+\frac{\left(2 L^2
   \rho _-+\rho _-+\rho _+\right) \log \rho _+}{8 L^2 \rho _++8 \rho
   _+}+\frac{1}{8} \left(\frac{\rho _-}{\rho _+}-1\right) \log ^2\rho _+
   \right).\label{eq:charged-kappa}
\end{align}\end{subequations}
And we must recall the potential on the horizon was fixed by the parametrization~(\ref{eq:charge-parametrization})
\begin{align}
 \Phi_H = \sqrt{\frac{\rho_-}{2\rho_+}}.
\end{align}
It is can be shown from direct computations that these quantities satisfy the first law and Smarr's relation
\begin{align}
& d{\cal M} = \kappa {\cal A}_H + \Phi_H dQ + {\cal T} dL,\\
 &2n {\cal M} = (2n+1) \kappa {\cal A}_H + {\cal T} L + 2n \Phi_H Q.
\end{align}
We omitted the expression for ${\cal M}_K$ as it follows Smarr's relation
\begin{align}
{\cal M}_K =  \frac{2n+1}{2n} \kappa {\cal A}_H + \Phi_H Q
 = {\cal M} -\frac{{\cal T}L}{2n}
\end{align}
The squashing function is written as
\begin{align}
&k_{\rm sq} = \frac{L^2}{1+L^2}\left[1-\frac{\log (\rho_+-\rho_-)}{\left(L^2+1\right) n}
+\frac{1}{6 \left(L^2+1\right)^2 n^2}\left(
-\frac{3 \rho_-(L^2+1)}{\rho_+-\rho_-}\right.\right.\nonum
&\left.\left.\qquad \qquad-3 L^2 \log ^2(\rho_+-\rho_-)+3 \log (\rho_+-\rho_-)+(\pi ^2-9) L^2
\right)\right].\label{eq:charged-ksq}
\end{align}
For the charged case, each physical quantity obeys the same scaling laws as for the neutral case under
\begin{align}
   \rho_\pm \to C \rho_\pm,\quad L \to C^\fr{2n} L.
\end{align}

\subsection{Extremal limit}
Until now, the extremal limit of charged black holes has been paid no attention to in the large $D$ effective theory.
Since the extremal limit leads to the divergent behavior in the $1/D$ correction, it has been simply considered as the breakdown of the $1/D$-expansion~\cite{Emparan:2016sjk}.
Our charged solution exhibits the same symptom as well. 
From eq.~(\ref{eq:charged-kappa}), the zero temperature limit $\kappa \to 0$ yields
\begin{align}
 \rho_+ =\widetilde{\rho}_-:= \left(1-\fr{2n}+\fr{4n^2}+\ord{n^{-3}}\right)\rho_- \simeq \frac{2n}{2n+1}\rho_-. \label{eq:extremal-rho+_0}
\end{align}
In this limit, the leading order solution~(\ref{eq:charged-metricsol-LO}) admits a degenerate horizon.
However, the sub-leading solution~(\ref{eq:app-sol-charge-nlo}) diverges for $\rho_+ \to \widetilde{\rho}_-$, from which one might conclude the breakdown of the $1/n$ expansion.
However, we can show that the divergent behavior can actually be tamed to obtain the regular extremal limit by introducing the extremal parameter
\begin{align}
\chi:= \left(1-\widetilde{\rho}_-/\rho_+\right)^\fr{2n}.
\end{align}
All the divergent terms in the thermodynamical variables~(\ref{eq:thermo-Q})
can be absorbed into the extremal parameter as follows
\begin{align}
&{\cal M}=r_H^{2n} \left[ \widetilde{\cal M}\left(\frac{r_H \chi}{L} \right) +(1-\chi^{2n})\widetilde{\cal M}_-\left(\frac{r_H \chi}{L} \right)\right],\nonum 
& Q = 2r_H^{2n}\sqrt{2(1-\chi^{2n})}\,\widetilde{Q}\left(\frac{r_H \chi}{L} \right),\quad {\cal T} = \chi^{2n-1} \widetilde{\cal T}\left(\frac{r_H \chi}{L} \right),\nonum
& \Phi_H = \sqrt{\frac{2n+1}{4n}(1-\chi^{2n})},\quad
 {\cal A}_H = r_H^{2n+1} \widetilde{\cal A}_H\left(\frac{r_H \chi}{L}\right),\quad
 \kappa = r_H^{-1} \chi^{2n} \widetilde{\kappa}\left(\frac{r_H \chi}{L}\right),
\end{align}
where $r_H:=\rho_+^\fr{2n}$ is the $CP^n$ radius on the horizon and
scale-invariant functions are the same as those of the neutral solution~(\ref{eq:neutral-scaleinv}) except the following two
\begin{align}
&\widetilde{\cal M}_-(x) =\frac{1}{\sqrt{1+x^2}}\left(1+\frac{3 x^2+2}{n \left(4
   x^2+4\right)}-\frac{\left(39 x^2-8 \pi ^2+72\right) x^2}{96 n^2 \left(x^2+1\right)^2}\right. \nonum
   &\qquad\left. -\frac{x^2 \left(x^2 \left(192 \zeta (3)+282-40 \pi ^2\right)+99 x^4+48 (8 \zeta (3)-9)\right)}{384 n^3
   \left(x^2+1\right)^3}\right),\\
&\widetilde{Q}(x) =\frac{1}{\sqrt{1+x^2}}\left(1+\frac{2 x^2+1}{n \left(4
   x^2+4\right)}+ \frac{-24 x^4+\left(8 \pi ^2-84\right) x^2-3}{96 n^2 \left(x^2+1\right)^2}\right.\nonum
&\qquad\left. +\, \frac{8 x^4 \left(4 \pi ^2-3 (8 \zeta (3)+9)\right)+x^2 \left(-384 \zeta (3)+522-8 \pi ^2\right)+3}{384   n^3 \left(x^2+1\right)^3}\right),
\end{align}
where NNNLO solutions are used to obtain $\ord{n^{-3}}$-correction.
The squashing function has exactly the same expression as in the neutral case~(\ref{eq:ksq-x})
\begin{align}
 k_{\rm sq} = \widetilde{k_{\rm sq}}\left(\frac{r_H \chi}{L}\right).
\end{align}
And hence, we obtain the regular extremal limit by $\chi \to 0$, in which the solution approaches that of the extremal Reissner-Nordstr\"{o}m black hole.\footnote{Here we do not show explicitly, but the metric solutions also have the smooth extremal limit, once the divergent terms are absorbed into $\chi$.} Particularly, the mass to charge ratio correctly saturates the known BPS bound~\cite{Tatsuoka:2011tx} up to $\ord{n^{-3}}$
\begin{align}
\frac{\cal M}{Q} = \fr{\sqrt{2}}\left(1+\fr{4n}-\fr{32n^2}+\fr{128n^3}\right)\approx \sqrt{\frac{2n+1}{4n}}.
\end{align}

\paragraph{Breakdown of $1/n$-expansion on the inner horizon}
In contrast to the extremal limit, the inner horizon seems to admit
a true breakdown of the $1/n$-expansion. In the leading order solution~(\ref{eq:charged-metricsol-LO}), assuming ${\sf R}-\rho_-={\cal O}(1/n)$, we have $B_0 \sim \ord{n}$ which breaks the expansion $B=1+B_0/n$. This means the near inner horizon region should be solved in a different setup at large $D$.

\section{Summary}\label{sec:sum}
In this paper, using the technique of the large $D$ effective theory,  we have studied the static solution of the extremal, neutral/charged SqKK black holes with 
the horizon topology of $S^{2n+1}$,  which we have viewed as an $S^1$ fiber bundle over $CP^n$ base.
First, we have solved the squashed horizonless background geometry by $1/D$ expansion which describes the generalized  Euclidean Taub-NUT spacetime with a flat timelike direction. 
Once the squashed background was solved in $1/D$-expansion,
the near horizon metric which asymptotes to the given background was obtained by the conventional large $D$ analysis. The neutral and charged solution were obtained almost in parallel. We evaluated the physical quantities for both cases. The extremal limit of the charged solution reproduced the consistent result with the known extremal analysis.

\medskip
In the analysis of the standard large $D$ effective theory, the main focus has been on the near-horizon geometry in the trivial background. We found that the large $D$ limit of the twisted background follows a non-trivial flow equation, which solves squashing in the twisted direction. This would open up a search for non-trivial backgrounds at large $D$. Interestingly, the conifold ansatz gives another first order flow equation, the Ricci flow, describing the topology-changing transition~\cite{Emparan:2019obu}. Although both solves first order flow equations, the squashing flow just solves the decoupled sector which does not involve the horizon, while the Ricci flow in the conifold ansatz can see the horizon which implies it involves the non-decoupled sector.

\medskip
The extremal limit and large $D$ limit have long been thought to be incompatible, since the large $D$ effective theory with a gauge field
admits sub-leading corrections divergent at the extremal limit.
Our result suggests that such divergent behavior merely reflects the existence of the extremal parameter in power of $1/D$, which simply requires a careful treatment at the large $D$ limit.
On the other hand, we found the divergent behavior on the inner horizon cannot be remedied by mere redefinition of parameters,  but requires another coordinate patch at large $D$. This could be related to the recent observation that the near extremal black hole at large $D$ exhibits two layers of the near horizon geometry which consists of $AdS_2$-throat and mid region.~\cite{Kachru:2020gmz}.

\medskip
We expect to be able to construct the rotating black hole solutions in the squashed background at large D, since in fact,  for five dimensions, the squashed rotating black holes  were obtained from asymptotically flat rotating  black holes with equal angular momenta via the squashing deformation~\cite{Dobiasch:1981vh,Nakagawa:2008rm}.
We also expect that there may be squashed black holes breaking the $U(1)$ symmetry along the fiber direction $\partial/\partial\phi$. To obtain such less symmetric solutions requires solving the large $D$ effective equation in the symmetry breaking direction.
Finally, we wish to comment on numerous generalizations of the Hopf fibration. 
 In particular, the Hopf fibration admits a family of  fiber bundles in which the total space, base space, and fiber space are all spheres, such as $S^7$ can be viewed as a fiber bundle of $S^3$ over $S^4$ base space. 
 It may be interesting to construct such squashed black hole solutions with nontrivial fiber. 
 The construction of these solutions deserves our future works.

\section*{Acknowledgement}
We thank Hideki Ishihara for useful comments on this project. RS was supported by JSPS KAKENHI Grant Number JP18K13541.
ST was supported by the Grant-in-Aid for Scientific Research (C) [JSPS KAKENHI Grant Number~17K05452] and the Grant-in-Aid for Scientific Research (C) [JSPS KAKENHI Grant Number~21K03560] from the Japan Society for the Promotion of Science. 

\appendix

\section{Reduction of spacetimes with an $S^1$ fiber over $CP^n$}\label{sec:A}
Here we show the reduction formula for a bundled spacetime used in the paper.
We consider the following spacetime ansatz with an $S^1$ fiber bundle over $CP^n$
\begin{eqnarray}
 ds^2 = G_{AB}(X)\xi^A \xi^B+r^2(X)\gamma_{ij}(\sigma)d\sigma^i d\sigma^j,
\end{eqnarray}
$\xi^A = dX^A+\delta^A_\phi \cA_n$ and $\cA_n = \cA_{n,i} d\sigma^i$ is the K\"{a}hlar potential on the $CP^n$ metric $\gamma_{ij}$. We assume the spacetime is symmetric in the bundled direction $\phi$.
To evaluate the Ricci tensor, we use the property of the K\"{a}hler form $J_{ij} = \partial_{[i} \cA_{n,j]}$
\begin{eqnarray}
\hat{\nabla}_i J{}_{jk}=0,\quad J{}_i{}^j J{}_{jk}=-\gamma_{ij},
\end{eqnarray}
where $\hat{\nabla}$ is the covariant derivative of $\gamma_{ij}$.
We obtain the decomposition of the Ricci tensor as
\begin{subequations}\label{eq:red-ricci}
\begin{align}
& R_{AB} = \bar{R}_{AB} -2n(r^{-1}\bar{\nabla}_A \bar{\nabla}_B r-r^{-4}G_{A\phi} G_{B\phi}),\label{eq:A3a}\\
& R_{Ai} = \cA_{n,i} R_{A\phi},\label{eq:A3b}\\
&R_{ij} =   \left[2(n+1)-2r^{-2}G_{\phi\phi}-(2n-1) (\bar{\nabla} r)^2-r\bar{\nabla}^2r  \right]\gamma_{ij}+R_{\phi\phi} \cA_{n,i} \cA_{n,j},\label{eq:A3c}
\end{align}
\end{subequations}
where we used the curvature of $CP^n$ given by $\hat{R}_{ij} = 2(n+1)\gamma_{ij}$. $\bar{\nabla}$ and $\bar{R}_{AB}$ represent the covariant derivative and Ricci tensor with respect to $G_{AB}$. In the mixed indices, we have

\begin{subequations}\label{eq:red-riccib}
\begin{align}
& R^A{}_B = \bar{R}^A{}_B -2n(r^{-1}\bar{\nabla}^A \bar{\nabla}_B r-r^{-4}\delta^A{}_\phi G_{B\phi}),\label{eq:A3a2}\\
& R^A{}_i = R^A{}_\phi \cA_{n,i}-\delta^A{}_\Phi \cA_{n,j} R^j{}_i,\quad R^i{}_A = 0,\\
&R^i{}_j =  \left[\frac{2(n+1)}{r^2}-\frac{2G_{\phi\phi}}{r^4}-(2n-1)\frac{ (\bar\nabla r)^2}{r^2}-\frac{\bar{\nabla}^2r}{r}  \right] \delta^i{}_j.
\end{align}
\end{subequations}

\section{Metric solutions in $1/D$ expansion}\label{sec:B}
Here, we present the metric solutions in $1/D$ expansion. Not to exhaust readers for the lengthly expression, some higher order solutions are provided in the auxiliary {\it Mathematica} notebook file in the supplementary material.
The metric ansatz are given by
\begin{align}
ds^2 = -A dt^2+h_{rr}\frac{B}{A}dr^2+h_{\phi\phi} H (d\phi+\cA_n)^2 + r^2 d\Sigma_n^2,
\end{align}
where $h_{rr}$ and $h_{\phi\phi}$ are the background Kaluza-Klein monopole solution~(\ref{eq:squashedbk-sol-bondi-LD}).
With the near horizon coordinate ${\sf R}:=r^{2n}$, the metric functions are expanded by $1/n$ as
\begin{align}
A = \sum_{i=0}^\infty \frac{A_i({\sf R})}{n^i},\quad B = 1 + \fr{n} \sum_{i=0}^\infty \frac{B_i({\sf R})}{n^i}
, \quad  H = 1 + \fr{n} \sum_{i=0}^\infty \frac{H_i({\sf R})}{n^i}.
\end{align}
In the charged solution, the gauge field $A_t=\Phi$ is also expanded by
\begin{equation}
 \Phi = \sum_{i=0}^\infty \frac{\Phi_i({\sf R})}{n^i}.
\end{equation}

\subsection{Neutral SqKK black holes}\label{app:sol-neutral}
First, we list the leading order $(A_0,B_0,H_0)$,  the next leading order $(A_1,B_1,H_1)$ and the next-to-next-to-leading order $(A_2,B_2,H_2)$ for the neutral squashed black holes.
\paragraph{Leading order}
\begin{align}
A_0 = 1-\frac{m}{\sf R} ,\quad B_0=0,\quad H_0=0.
\end{align}

\paragraph{Next-leading order}
\begin{align}
&A_1=-\frac{m \log \left(\frac{\sR}{m}\right)}{2 (L^2+1) \sR},\quad B_1 =- \frac{L^2 m \log(\frac{\sR}{m})}{(1 + L^2)^2 ( \sR-m)},\quad H_1 = \frac{L^2 \left(6\, \text{Li}_2\left(1-\frac{\sR}{m}\right)+3 \log ^2\left(\frac{\sR}{m}\right)+\pi
   ^2\right)}{6 \left(L^2+1\right)^2}.
\end{align}
\paragraph{Next-to-next-to-leading order}
\begin{align}
&A_2 = \frac{4 m L^2 {\rm Li}_2(1-\sR/m)-m\log (\sR/m) \left(4 L^2 \log m-2 L^2+\log (\sR/m)\right)}{8
   \left(L^2+1\right)^2 R},\nonum
 &B_2=\frac{L^2 \left(3 \left(L^2-1\right) m^2+\left(2-6 L^2\right) m \sR+\left(3 L^2-1\right) \sR^2\right) \log^2\left(\sR/m\right)}{4 \left(L^2+1\right)^3 (m-\sR)^2}\nonum
 &+\, \frac{\left(3 L^2-1\right) L^2  }{2 \left(L^2+1\right)^3}\text{Li}_2\left(1-\frac{\sR}{m}\right)
   +\frac{m \left(2 L^2
   \left(L^2-1\right) \log (m)-L^4\right) \log \left(\frac{\sR}{m}\right)}{2 \left(L^2+1\right)^3
   (m-\sR)}+\frac{\pi ^2 \left(3 L^2-1\right) L^2}{12 \left(L^2+1\right)^3},\nonum
 &H_2=\frac{\left(2 L^2+1\right)L^2}{\left(L^2+1\right)^3}\left(\text{Li}_3\left(1-\frac{m}{\sR}\right)+ \text{Li}_3\left(1-\frac{\sR}{m}\right)\right)
-\frac{\left(L^2+2\right) L^2 \log
   ^3\left(\frac{\sR}{m}\right)}{6 \left(L^2+1\right)^3}
   \nonum
   &+\text{Li}_2\left(1-\frac{\sR}{m}\right)
   \left(\frac{L^2 \left(\left(L^2-1\right) \log (m)+1\right)}{\left(L^2+1\right)^3}-\frac{L^2 \log
   \left(\frac{\sR}{m}\right)}{\left(L^2+1\right)^2}\right)  +\frac{\pi ^2 L^4  \log \left(\frac{\sR}{m}\right)}{6 \left(L^2+1\right)^3} \nonum
&  -\frac{L^2 \log ^2\left(\frac{\sR}{m}\right)   \left(\sR-\left(L^2-1\right) (m-\sR) \log m\right)}{2 \left(L^2+1\right)^3 (m-\sR)}
   +\frac{L^2 \left(\pi ^2 \left(L^2-1\right)
   \log (m)-6 \left(2 L^2+1\right) \zeta (3)+\pi ^2\right)}{6 \left(L^2+1\right)^3}.
\end{align}
NNNLO solutions are presented in the auxiliary file in the supplementary material..

\subsection{Charged SqKK black holes}\label{app:sol-charged}

Next, we list  the leading order $(A_0,B_0,H_0,\Phi_0)$ and  the next leading order $(A_1,B_1,H_1,\Phi_1)$
for the charged squashed black holes.
\paragraph{Leading order}
\begin{align}
& \Phi_0 = \frac{ \sqrt{\rho_+ \rho_-}}{\sqrt{2} {\sf R}},\quad
 A_0 = 1-\frac{\rho_++\rho_-}{{\sf R}}+\frac{\rho_+\rho_-}{{\sf R}^2},\nonum
 &H_0 = \frac{1}{1+L^2}\log \left(1-\frac{\rho_-}{\sf R}\right),\quad
  B_0 = -\frac{\rho_-}{(1+L^2)({\sf R}-\rho_-)}.
\end{align}

\paragraph{Next-to-leading order}
\begin{align}
& \Phi_1 = \frac{\sqrt{\rho_+\rho_-}}{2\sqrt{2}(1+L^2){\sf R}}\log\left(\frac{{\sf R}-\rho_-}{\rho_+-\rho_-} \right),\nonum
& A_1 = -\frac{\rho_-({\sf R}-\rho_+)}{2{\sf R}^2}
- \frac{(\rho_++\rho_-){\sf R}-2\rho_+\rho_-}{2(1+L^2){\sf R}^2}\log \left(\frac{{\sf R}-\rho_-}{\rho_+-\rho_-}\right),\nonumber
\end{align}
\begin{align}
& H_1 = \frac{L^2}{\left(L^2+1\right)^2} \text{Li}_2\left(\frac{\rho _+-\sR}{\rho _+-\rho   _-}\right)+\frac{\log ^2\left(\sR-\rho _-\right)}{2   \left(L^2+1\right)^2}+\frac{\log ^2\sR}{2 (L^2+1)} -\frac{\log \sR \log \left(\sR-\rho   _-\right)}{\left(L^2+1\right)^2}\nonum
   &+\frac{\left(\left(3+\pi ^2\right) L^2+3\right) \rho _-+3 L^2   \log ^2\left(\rho _+-\rho _-\right) \left(\rho _- -\sR\right)-\pi ^2 L^2 \sR+3 \rho _-   \log \left(\rho _+-\rho _-\right)}{6 \left(L^2+1\right)^2 \left(\rho   _--\sR\right)}\nonum
&  -\frac{\log \left(\sR-\rho _-\right) \left(2 L^2
   \log \left(\rho _+-\rho _-\right) \left(\rho _--\sR\right)+\rho _-\right)}{2
   \left(L^2+1\right)^2 \left(\rho _--\sR\right)},\nonum
& B_1=-\left(\frac{\rho _+ \left(2 L^2 \rho _-^2+2 L^2 \sR^2-\rho _- \left(4 L^2
   \sR+\sR\right)\right)+\rho _- \sR^2}{2\left(L^2+1\right)^2\left(\rho _--\sR\right){}^2 \left(\sR-\rho
   _+\right)}+\frac{3 L^2}{2 \left(L^2+1\right)^2}\right) \log \left(\sR-\rho _-\right)\nonum
&   -\frac{L^2 \rho _+ \log \left(\rho _+-\rho   _-\right)}{\left(L^2+1\right)^2 \left(\rho _+-\sR\right)}-\frac{\rho _-   \left(\left(1-2 L^2\right) \log \left(\rho _+-\rho _-\right)+2 L^2+1\right)}{2
   \left(L^2+1\right)^2 \left(\rho _--\sR\right)}\nonum
   &+\frac{\rho _-^2 \left(2 L^2+2 \log
   \left(\rho _+-\rho _-\right)+5\right)}{4 \left(L^2+1\right)^2 \left(\rho
   _--\sR\right){}^2}+\frac{3 L^2 \log \sR}{2 \left(L^2+1\right)^2}.   \label{eq:app-sol-charge-nlo}
\end{align}
NNLO and part of NNNLO ($\Phi_3,\, A_3$) are presented in the auxiliary file in the supplementary material..


\begin{thebibliography}{99}
\bibitem{Emparan:2008eg}
R.~Emparan and H.~S.~Reall,
``Black Holes in Higher Dimensions,''
Living Rev. Rel. \textbf{11}, 6 (2008)
[arXiv:0801.3471 [hep-th]].



\bibitem{Kaluza:1921tu}
T.~Kaluza,``Zum Unit\"atsproblem der Physik,''
Sitzungsber. Preuss. Akad. Wiss. Berlin (Math. Phys. ) \textbf{1921}, 966-972 (1921)
[arXiv:1803.08616 [physics.hist-ph]].



\bibitem{Klein:1926tv}
O.~Klein,
``Quantum Theory and Five-Dimensional Theory of Relativity. (In German and English),''
Z. Phys. \textbf{37}, 895-906 (1926)


\bibitem{Dobiasch:1981vh}
P.~Dobiasch and D.~Maison,
``Stationary, Spherically Symmetric Solutions of Jordan's Unified Theory of Gravity and Electromagnetism,''
Gen. Rel. Grav. \textbf{14}, 231-242 (1982)


\bibitem{Ishihara:2005dp}
H.~Ishihara and K.~Matsuno,
``Kaluza-Klein black holes with squashed horizons,''
Prog. Theor. Phys. \textbf{116}, 417-422 (2006)
[arXiv:hep-th/0510094 [hep-th]].

\bibitem{Gibbons:1985ac}
G.~W.~Gibbons and D.~L.~Wiltshire,``Black Holes in Kaluza-Klein Theory,''
Annals Phys. \textbf{167}, 201-223 (1986)
[erratum: Annals Phys. \textbf{176}, 393 (1987)]
doi:10.1016/S0003-4916(86)80012-4




\bibitem{Nakagawa:2008rm}
T.~Nakagawa, H.~Ishihara, K.~Matsuno and S.~Tomizawa,
``Charged Rotating Kaluza-Klein Black Holes in Five Dimensions,''
Phys. Rev. D \textbf{77}, 044040 (2008)
[arXiv:0801.0164 [hep-th]].


\bibitem{Tomizawa:2008hw}
S.~Tomizawa, H.~Ishihara, K.~Matsuno and T.~Nakagawa,
``Squashed Kerr-Godel Black Holes: Kaluza-Klein Black Holes with Rotations of Black Hole and Universe,''
Prog. Theor. Phys. \textbf{121}, 823-841 (2009)
[arXiv:0803.3873 [hep-th]].



\bibitem{Tomizawa:2008rh}
S.~Tomizawa and A.~Ishibashi,
``Charged Black Holes in a Rotating Gross-Perry-Sorkin Monopole Background,''
Class. Quant. Grav. \textbf{25}, 245007 (2008)
[arXiv:0807.1564 [hep-th]].

\bibitem{Tatsuoka:2011tx}
T.~Tatsuoka, H.~Ishihara, M.~Kimura and K.~Matsuno,
``Extremal Charged Black Holes with a Twisted Extra Dimension,''
Phys. Rev. D \textbf{85}, 044006 (2012)
[arXiv:1110.6731 [hep-th]].




\bibitem{Emparan:2013moa}
  R.~Emparan, R.~Suzuki and K.~Tanabe,
  ``The large D limit of General Relativity,''
  JHEP {\bf 1306} (2013) 009
  [arXiv:1302.6382 [hep-th]].
  

\bibitem{Emparan:2020vyfinr}
R.~Emparan and C.~P.~Herzog,
``The Large D Limit of Einstein's Equations,''
[arXiv:2003.11394 [hep-th]].


\bibitem{Emparan:2015hwa}
R.~Emparan, T.~Shiromizu, R.~Suzuki, K.~Tanabe and T.~Tanaka,
``Effective theory of Black Holes in the 1/D expansion,''
JHEP \textbf{06} (2015), 159
[arXiv:1504.06489 [hep-th]].


\bibitem{Bhattacharyya:2015dva}
S.~Bhattacharyya, A.~De, S.~Minwalla, R.~Mohan and A.~Saha,
``A membrane paradigm at large D,''
JHEP \textbf{04} (2016), 076
[arXiv:1504.06613 [hep-th]].

\bibitem{Bhattacharyya:2015fdk}
S.~Bhattacharyya, M.~Mandlik, S.~Minwalla and S.~Thakur,
``A Charged Membrane Paradigm at Large D,''
JHEP \textbf{04} (2016), 128
[arXiv:1511.03432 [hep-th]].

\bibitem{Suzuki:2015axa}
R.~Suzuki and K.~Tanabe,
``Non-uniform black strings and the critical dimension in the $1/D$ expansion,''
JHEP \textbf{10}, 107 (2015)
[arXiv:1506.01890 [hep-th]].

\bibitem{Emparan:2015gva}
  R.~Emparan, R.~Suzuki and K.~Tanabe,
  ``Evolution and End Point of the Black String Instability: Large D Solution,''
  Phys.\ Rev.\ Lett.\  {\bf 115} (2015) no.9,  091102
  [arXiv:1506.06772 [hep-th]].



\bibitem{Emparan:2016sjk}
  R.~Emparan, K.~Izumi, R.~Luna, R.~Suzuki and K.~Tanabe,
  ``Hydro-elastic Complementarity in Black Branes at large D,''
  JHEP {\bf 1606} (2016) 117
  [arXiv:1602.05752 [hep-th]].
  
  \bibitem{Rozali:2016yhw}
M.~Rozali and A.~Vincart-Emard,
``On Brane Instabilities in the Large $D$ Limit,''
JHEP \textbf{08}, 166 (2016)
[arXiv:1607.01747 [hep-th]].
  
\bibitem{Emparan:2018bmi}
R.~Emparan, R.~Luna, M.~Mart\'{i}nez, R.~Suzuki and K.~Tanabe,
``Phases and Stability of Non-Uniform Black Strings,''
JHEP \textbf{05} (2018), 104
[arXiv:1802.08191 [hep-th]].






\bibitem{Tanabe:2015hda}
K.~Tanabe,
``Black rings at large D,''
JHEP \textbf{02}, 151 (2016)
doi:10.1007/JHEP02(2016)151
[arXiv:1510.02200 [hep-th]].

\bibitem{Tanabe:2016opw}
K.~Tanabe,
``Charged rotating black holes at large D,''
[arXiv:1605.08854 [hep-th]].

\bibitem{Chen:2017wpf}
B.~Chen, P.~C.~Li and Z.~z.~Wang,
``Charged Black Rings at large D,''
JHEP \textbf{04}, 167 (2017)
[arXiv:1702.00886 [hep-th]].

\bibitem{Mandlik:2018wnw}
M.~Mandlik and S.~Thakur,
``Stationary Solutions from the Large D Membrane Paradigm,''
JHEP \textbf{11}, 026 (2018)
[arXiv:1806.04637 [hep-th]].





\bibitem{Iizuka:2018zgt}
N.~Iizuka, A.~Ishibashi and K.~Maeda,
``Cosmic Censorship at Large D: Stability analysis in polarized AdS black branes (holes),''
JHEP \textbf{03}, 177 (2018)
[arXiv:1801.07268 [hep-th]].



\bibitem{Herzog:2017qwp}
C.~P.~Herzog and Y.~Kim,
``The Large Dimension Limit of a Small Black Hole Instability in Anti-de Sitter Space,''
JHEP \textbf{02}, 167 (2018)
[arXiv:1711.04865 [hep-th]].







\bibitem{Andrade:2018nsz}
  T.~Andrade, R.~Emparan and D.~Licht,
  ``Rotating black holes and black bars at large D,''
  JHEP {\bf 1809} (2018) 107
  [arXiv:1807.01131 [hep-th]].


\bibitem{Andrade:2018rcx}
  T.~Andrade, R.~Emparan and D.~Licht,
  ``Charged rotating black holes in higher dimensions,''
  JHEP {\bf 1902} (2019) 076
  [arXiv:1810.06993 [hep-th]].

\bibitem{Andrade:2018yqu}
  T.~Andrade, R.~Emparan, D.~Licht and R.~Luna,
  ``Cosmic censorship violation in black hole collisions in higher dimensions,''
  JHEP {\bf 1904} (2019) 121
  [arXiv:1812.05017 [hep-th]].


\bibitem{Andrade:2019edf}
  T.~Andrade, R.~Emparan, D.~Licht and R.~Luna,
  ``Black hole collisions, instabilities, and cosmic censorship violation at large $D$,''
  JHEP {\bf 1909} (2019) 099
  [arXiv:1908.03424 [hep-th]].
 
\bibitem{Licht:2020odx}
D.~Licht, R.~Luna and R.~Suzuki,
``Black Ripples, Flowers and Dumbbells at large $D$,''
JHEP \textbf{04}, 108 (2020)
[arXiv:2002.07813 [hep-th]].

\bibitem{Andrade:2020ilm}
T.~Andrade, R.~Emparan, A.~Jansen, D.~Licht, R.~Luna and R.~Suzuki,
``Entropy production and entropic attractors in black hole fusion and fission,''
JHEP \textbf{08}, 098 (2020)
[arXiv:2005.14498 [hep-th]].

\bibitem{Suzuki:2020kpx}
R.~Suzuki,
``Black hole interactions at large $D$: brane blobology,''
JHEP \textbf{02}, 131 (2021)
[arXiv:2009.11823 [gr-qc]].





\bibitem{Chen:2017rxa}
B.~Chen, P.~C.~Li and C.~Y.~Zhang,
``Einstein-Gauss-Bonnet Black Strings at Large $D$,''
JHEP \textbf{10}, 123 (2017)
[arXiv:1707.09766 [hep-th]].

\bibitem{Chen:2018vbv}
B.~Chen, P.~C.~Li and C.~Y.~Zhang,
``Einstein-Gauss-Bonnet Black Rings at Large $D$,''
JHEP \textbf{07}, 067 (2018)
[arXiv:1805.03345 [hep-th]].




\bibitem{Hoxha:2000jf}
P.~Hoxha, R.~R.~Martinez-Acosta and C.~N.~Pope,
``Kaluza-Klein consistency, Killing vectors, and Kahler spaces,''
Class. Quant. Grav. \textbf{17}, 4207-4240 (2000)
[arXiv:hep-th/0005172 [hep-th]].



\bibitem{Harmark:2004ch}
T.~Harmark and N.~A.~Obers,
``General definition of gravitational tension,''
JHEP \textbf{05}, 043 (2004)
[arXiv:hep-th/0403103 [hep-th]].





\bibitem{Emparan:2019obu}
R.~Emparan and R.~Suzuki,
``Topology-changing horizons at large D as Ricci flows,''
JHEP \textbf{07}, 094 (2019)
[arXiv:1905.01062 [hep-th]].

\bibitem{Kachru:2020gmz}
S.~Kachru and M.~Shyani,
``Holographic non-Fermi liquids at large $d$,''
[arXiv:2010.03560 [hep-th]].

\bibitem{Kurita:2007hu}
Y.~Kurita and H.~Ishihara,
``Mass and free energy in thermodynamics of squashed Kaluza-Klein black holes,''
Class. Quant. Grav. \textbf{24}, 4525-4532 (2007)
[arXiv:0705.0307 [hep-th]].


\end{thebibliography}
\end{document}